\documentclass[]{article}
\usepackage{times}

\usepackage{amssymb}
\usepackage[hang,scriptsize]{subfigure}
\usepackage{latexsym}
\usepackage{lastpage}
\usepackage{fancyhdr}

\newif\ifpdf
\ifx\pdfoutput\undefined
   \pdffalse
\else
   \pdfoutput=1
   \pdftrue
\fi
\ifpdf
  \usepackage[pdftex]{graphicx}
 
\usepackage[pdftex,colorlinks=true,pdfstartview=FitH,linkcolor=black,citecolor=black,urlcolor=black]{hyperref}
\else
  \usepackage{graphicx}

  \newcommand{\href}[2]{#2}
  
\fi

\usepackage{boxedminipage}

\makeatletter
\def\@maketitle{%
  \vbox to 2.3in{%
    \hsize\textwidth
    \linewidth\hsize
    \vspace*{1.5cm}
    \centering
    {\bfseries\huge \@title \par}
    \vskip 2em
    {\large \begin{tabular}[t]{c}\@author \end{tabular}\par}
    \vfill}    \vspace*{1.0cm}
}
\renewcommand\section{\@startsection {section}{1}{\z@}%
     {.7\baselineskip plus\baselineskip}{.5\baselineskip}
                                   {\normalfont\Large\bfseries}}
\renewcommand\section{\@startsection {section}{1}{\z@}%
      {.5\baselineskip\@plus.7\baselineskip}{.3\baselineskip}%
                                   {\normalfont\Large\bfseries}}
\renewcommand\subsection{\@startsection{subsection}{2}{\z@}%
       {.5\baselineskip\@plus.7\baselineskip}{.3\baselineskip}%
                                   {\normalfont\large\bfseries}}
\renewcommand\subsubsection{\@startsection{subsubsection}{3}{\z@}%
      {.5\baselineskip\@plus.7\baselineskip}{.3\baselineskip}%
                                     {\normalfont\normalsize\bfseries}}
\makeatother
 
\renewenvironment{abstract}%
  {\normalfont
    \list{}{\labelwidth0pt
      \leftmargin0pt \rightmargin\leftmargin
      \listparindent\parindent \itemindent0pt
      \parsep0pt
      
    }%
    \item[\hskip\labelsep\bfseries\abstractname\enspace --] \itshape%
}{%
  \endlist}

\newcommand{\keywordsname}{Keywords}
\newenvironment{keywords}%
  {\normalfont
    \list{}{\labelwidth0pt
      \leftmargin0pt \rightmargin\leftmargin
      \listparindent\parindent \itemindent0pt
      \parsep0pt
      }%
    \item[\hskip\labelsep\bfseries\keywordsname:]}{\endlist}

\newcommand{\bi}{\begin{itemize}}
\newcommand{\ei}{\end{itemize}}
\newcommand{\be}{\begin{equation}}
\newcommand{\ee}{\end{equation}}
\newcommand{\bea}{\begin{eqnarray}}
\newcommand{\eea}{\end{eqnarray}}
\newcommand{\ba}{\begin{array}}
\newcommand{\ea}{\end{array}}

\begin{document}
\title{Goodness-of-fit of the Heston model}
\author{
Gilles Daniel$^1$ ~~~ David S. Br\'ee$^1$ ~~~ Nathan L. Joseph$^2$\\[0.3cm]
$^1$Computer Science Department\\
$^2$Manchester School of Accounting \& Finance\\ 
University of Manchester, U.K.\\
{ \tt gilles.daniel@cs.man.ac.uk}\\
{ \tt http://www.cs.man.ac.uk/\~{}danielg}
}
%\date{}

\maketitle
\thispagestyle{empty}
\begin{abstract}

An analytical formula for the probability distribution of stock-market returns, derived from the Heston model assuming a mean-reverting stochastic volatility, was recently proposed by Dr\u{a}gulescu and Yakovenko in Quantitative Finance 2002. While replicating their results, we found two significant weaknesses in their method to pre-process the data, which cast a shadow over the effective goodness-of-fit of the model. We propose a new method, more truly capturing the market, and perform a Kolmogorov-Smirnov test and a $\chi^2$ test on the resulting probability distribution. The results raise some significant questions for large time lags --- $40$ to $250$ days --- where the smoothness of the data does not require such a complex model; nevertheless, we also provide some statistical evidence in favour of the Heston model for small time lags --- $1$ and $5$ days --- compared with the traditional Gaussian model assuming constant volatility.

\end{abstract}


\section{Assessed model}
\label{DYformula}

Standard models of stock-market fluctuations predict a normal (Gaussian) distribution for stock price log-returns \cite{Wilmott}.  However, the empirical distribution exhibits significant kurtosis with a greater probability mass in its tails and centre of the distribution \cite{Mandelbrot}. In Quantitative Finance 2002 (\cite{DY}), Dr\u{a}gulescu and Yakovenko (DY) proposed an analytical formula for the probability density function (\emph{pdf}) of stock price log-returns, based on the Heston model \cite{Heston} --- a geometric Brownian motion for the log-returns time series coupled with a stochastic mean-reverting volatility. The resulting \emph{pdf} is claimed to outperform the Gaussian on a large scale of time lags ($t = 1, 5, 20, 40$ and $250$ days).\\

\subsection{Heston model and DY formula}

This model starts with the usual assumption that the price $S_t$ follows a geometric Brownian motion described by the following stochastic differential equation
\begin{equation} \label{eqn:stochPrice}
dS_t = \mu S_t dt + \sigma_t S_t dW_t^{(1)}  
\end{equation}
\begin{tabular}{ll}
where & $\mu$ is the trend of the market,\\
	  & $\sigma_t$ is the volatility,\\
	  & $W_t^{(1)}$ is a standard Wiener process.\\
\end{tabular}\\
Log-returns $r_t=Log \frac{S_t}{S_0}$ and centred log-returns $x_t = r_t - \mu t$ are then introduced:
\begin{equation} \label{eqn:stochLogReturns1}
dr_t = (\mu-\frac{v_t}{2}) dt + \sqrt{v_t} dW_t^{(1)} \ \ \ \mbox{since} \ \sigma_t=\sqrt{v_t}
\end{equation}
and
\begin{equation} \label{eqn:stochLogReturns}
dx_t = -\frac{v_t}{2} dt + \sqrt{v_t} dW_t^{(1)}  
\end{equation}
Then, instead of having a constant volatility $\sigma_t=\sigma$ as in the Bachelier-Osborne model \cite{Bachelier, Osborne}, the Heston model assumes the variance $v_t=\sigma_t^2$ obeys the following mean-reverting stochastic differential equation
\begin{equation} \label{eqn:stochVar}
dv_t = -\gamma (v_t - \theta) dt + k \sqrt{v_t} dW_t^{(2)} 
\end{equation}
\begin{tabular}{ll} 
where & $\theta$ is the long time mean of $v_t$, \\
      & $\gamma$ is the relaxation rate of this mean, \\
      & $k$ is a constant parameter called the variance noise, \\
      & $dW_t^{(2)}$ is another standard Wiener process, not necessarily correlated with $dW_t^{(1)}$. \\
\end{tabular}\\
DY solve the forward Kolmogorov equation that governs the time evolution of the joint probability $P_t(x,v|v_i)$ of having the log-return $x$ and the variance $v$ for the time lag $t$, given the initial value $v_i$ of the variance
\begin{eqnarray}\label{eqn:forwardKolmogorov}
\frac{\delta}{\delta t} P & = & \gamma \frac{\delta}{\delta v} [(v-\theta)P] + \frac{1}{2} \frac{\delta}{\delta x} (vP) \nonumber \\
&  & + \rho k \frac{\delta^2}{\delta x \delta v} (vP) + \frac{1}{2} \frac{\delta^2}{\delta x^2} (vP) + \frac{k^2}{2} \frac{\delta^2}{\delta v^2} (vP)
\end{eqnarray}
They introduce a Fourier transform to solve analytically this equation, and obtain the following expression for the probability distribution of centred log-returns $x$, given a time lag $t$:
\begin{equation} \label{eqn:draguPDF}
P_t(x) = \frac{1}{2 \pi} \int_{-\infty}^{+\infty}{dp_x e^{ip_x+F_t(p_x)}}
\end{equation}
with
\begin{equation} \label{eqn:Ft}
F_t(p_x) = \frac{\gamma \Gamma \theta t}{k^2}
- \frac{2\gamma\theta}{k^2}ln[\frac{cosh\frac{\Omega t}{2} + \frac{\gamma}{\Omega}sinh\frac{\Omega t}{2}}{cosh\frac{\Omega t}{2} + \frac{\Gamma}{\Omega}sinh\frac{\Omega t}{2}}] 
- \frac{2 \gamma \theta}{k^2}ln[cosh\frac{\Omega t}{2} + \frac{\Omega ^2 - \Gamma ^2 + 2\gamma \Gamma}{2 \gamma \Omega}sinh\frac{\Omega t}{2}] 
\end{equation}
\begin{tabular}{ll}
where & $\Gamma = \gamma + i \rho k p_x$, \\
      & $\rho$ is the correlation coefficient between the two Wieners $W_t^{(1)}$ and $W_t^{(2)}$, \\
      & $\Omega = \sqrt{\Gamma^2 + k^2 (p_x^2 - ip_x)}$, \\
      & $\gamma$, $\theta$, $k$ and $\mu$ are the parameters of the Heston model. \\
\end{tabular}\\
Eqn. (\ref{eqn:draguPDF}), hereafter designated as the ``DY formula'', is the central result of DY's paper \cite{DY}. It gives the expected probability distribution of centred log-returns $x$, given a time lag $t$. An asymptotic analysis\footnote{See \cite{DY} Part VI} of $P_t(x)$ shows that it predicts a Gaussian distribution for small values of $|x|$, and exponential, time dependent tails for large values of $|x|$.\\

To confront their formula with observed log-returns, DY take the Dow-Jones Industrial Average from January 04, 1982 to December 31, 2001, and train the four parameters of the Heston model, $\gamma$, $\theta$, $k$ and $\mu$, to fit the empirical distribution by minimising the following square-mean deviation error 
$$
E = \sum_{x,t} |log P_t^*(x) - log P_t(x)|
$$
for all available values of log-returns $x$, and time lags $t = 1, 5, 20, 40$ and $250$ days, where $P_t^*(x)$ is the empirical probability mass and $P_t(x)$ the one predicted by the DY formula. In their results, they set the correlation coefficient $\rho$ to zero, since (i) their trained parameter $\rho^{trained}$ is almost null ($\rho^{trained} \simeq 0$), and (ii) they do not observe any significant difference, when fitting empirical data, between taking $\rho^{trained}\simeq 0$ or $\rho = 0$. Hence, they reduce the complexity of their formula. Minimising the deviation of the log instead of the absolute difference $|P_t^*(x) - P_t(x)|$ forces the parameters to fit the fat tails instead of the middle of the distribution, where the probability mass is very high.\\

\begin{figure}[t]
\centering
\vspace{2.5mm}
\centerline{\includegraphics[width=.75 \columnwidth]{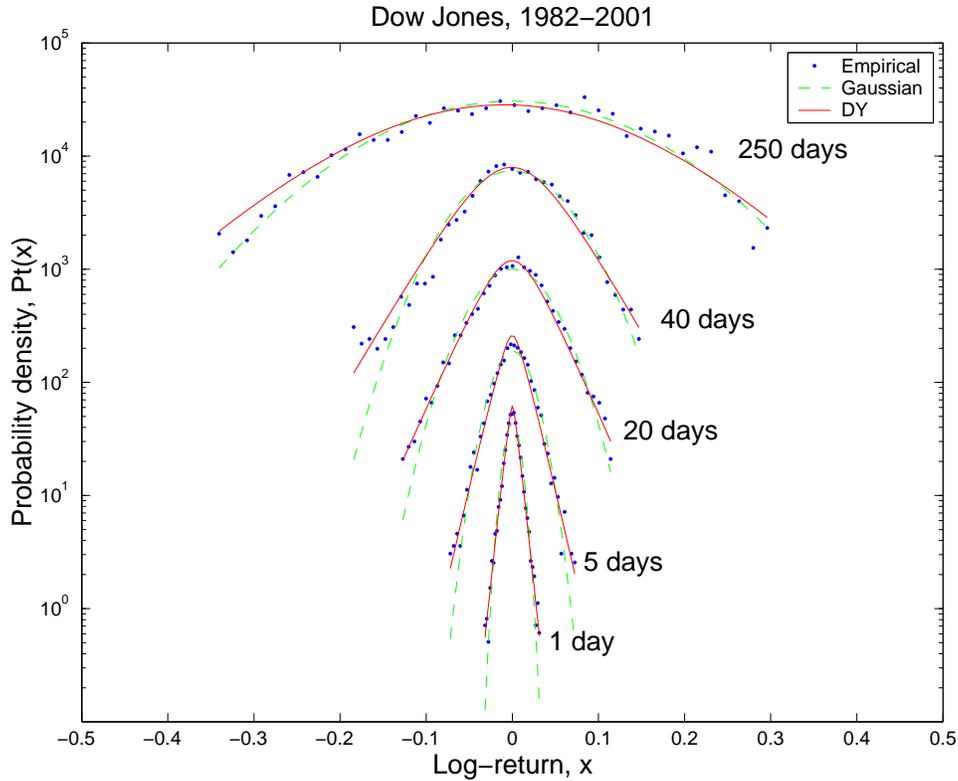}}
\vspace{2mm}
\caption{Replication of the main result of DY, where outliers are trimmed and data are reused.}
\label{fig:fig-DJIA8201-Trim-Reuse}
\end{figure}

A replication of their results, using the same dataset and the same method, is shown in Figure \ref{fig:fig-DJIA8201-Trim-Reuse}. The \emph{pdfs} for different time lags from 1 to 250   are shown each shifted upwards by a factor of 10 for clarity. Theoretically, this result is brilliant, since the authors obtain an analytic expression for the \emph{pdf} $P_t(x)$. Furthermore, their model (plain line) seems to fit the empirical data (dots) far better than the Gaussian (dash line), especially if we look at the fat tails.\\

But we argue that the method they use to evaluate the goodness-of-fit of their model suffers from two major drawbacks when pre-processing the data. 

\subsection{Pre-processing the data}

Firstly, DY trimmed the log-returns time series, rejecting any value out of the boundaries presented in Table \ref{tab:boundTrim}.\footnote{This step is not mentioned in DY's paper. Before applying this trimming method, strange points used to appear in our results. Then we contacted the authors who informed us they had trimmed the log-returns time series using the boundaries specified in Table \ref{tab:boundTrim}.} By doing so, they remove most of the leptokurtosis of the original dataset (the positive peakiness of the \emph{pdf} in the centre), which is precisely one of the discrepancies from the Gaussian that they should try to fit. 

\begin{table}[h]
\begin{center}
\begin{tabular}{cc}
time lag  & trimming boundaries \\ 
\begin{tabular}{r}
	   1 \\
	   5 \\
	  20 \\
	  40 \\
	  80 \\
	 100 \\
	 200 \\
	 250 \\
	\end{tabular} &
\begin{tabular}{c}
	$[-0.04 \ 0.04]$ \\
	$[-0.08 \ 0.08]$ \\
	$[-0.13 \ 0.15]$ \\
	$[-0.17 \ 0.20]$ \\
	$[-0.18 \ 0.25]$ \\
	$[-0.20 \ 0.28]$ \\
	$[-0.22 \ 0.38]$ \\
	$[-0.22 \ 0.44]$ \\
	\end{tabular} \\
\end{tabular}
\end{center}
\caption{Boundaries used by DY to trim the empirical log-returns time series.}
\label{tab:boundTrim}
\end{table}

\begin{figure}[t]
\centering
\vspace{2.5mm}
\centerline{\includegraphics[width=.75 \columnwidth]{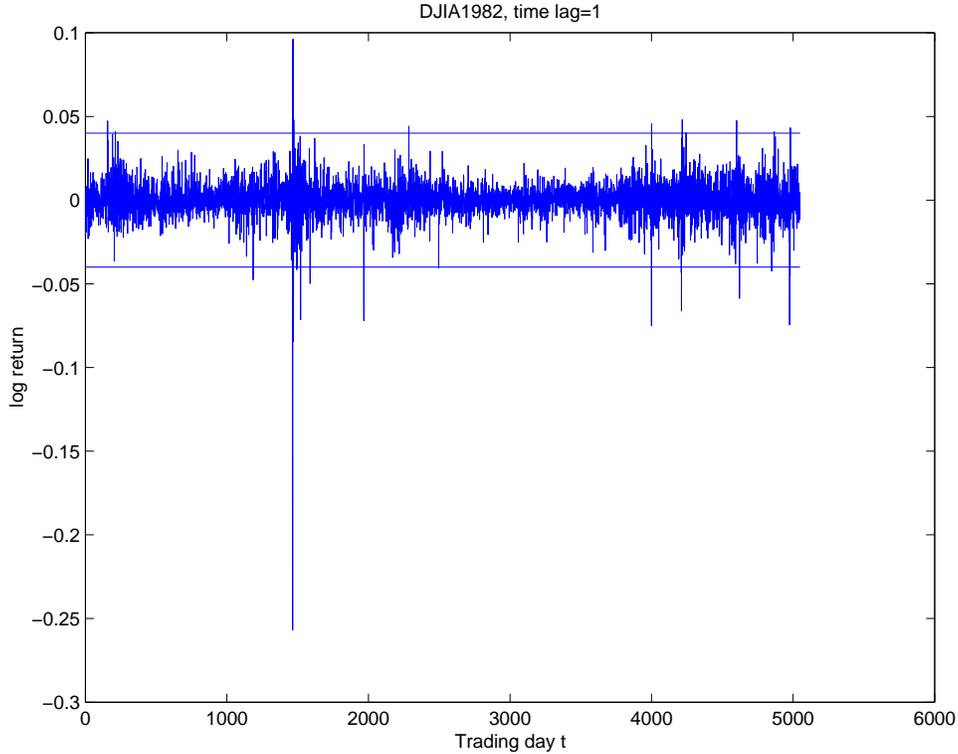}}
\vspace{2mm}
\caption{Boundaries used by DY to trim the log-returns, for daily returns.}
\label{fig:fig-trimmedLogReturns}
\end{figure}

We visualise in Figure \ref{fig:fig-trimmedLogReturns} the effect of trimming the data: all of the log-returns outside the boundaries, represented here by the two horizontal lines, were trimmed. We believe this way of trimming the data is unfair, because it removes information from the dataset. Given that the model is supposed to outperform the Bachelier-Osborne model, and specially to fit the kurtosis and the fat tails, removing extreme values (that belong to the fat tails, and produce kurtosis) is counter-productive. Even the normal distribution could fit the data quite well in these conditions \cite{GD}. To prove this, we compare the kurtosis of trimmed and untrimmed empirical data in Table \ref{tab:kurtosis}.

\begin{table}[h]
\begin{center}
\begin{tabular}{ccc}
time lag  & $k_{untrimmed}$ & $k_{trimmed}$ \\ 
\begin{tabular}{r}
	   1 \\
	   5 \\
	  20 \\
	  40 \\
	 250 \\
	\end{tabular} &
\begin{tabular}{r@{}l}
	69&.27 \\
	19&.68 \\
	7&.80 \\
	6&.02 \\
	-0&.33 \\
	\end{tabular} &
\begin{tabular}{r@{}l}
	1&.40 \\
	0&.72 \\
	0&.43 \\
	0&.56 \\
	-0&.53 \\
	\end{tabular} \\
\end{tabular}
\end{center}
\caption{Comparison of the kurtosis of trimmed and untrimmed empirical data.}
\label{tab:kurtosis}
\end{table}

Obviously, the kurtosis disappears when data are trimmed, which makes the dataset considerably smoother. Back in 1965, Fama \cite{Fama} had already made similar criticism of Kendall's experiments \cite{Kendal}, in which the latter considered outliers so extreme that he just dropped them.

A second major concern is that DY use a single log-return time series of overlapping returns. For a given index \emph{I} at a given period, let us say the Dow Jones Industrial Average from January 04, 1982, to December 31, 2001, and a given time lag $\tau$, let us say $\tau = 5$ days, the raw close price dataset \emph{closePrice} is composed of \emph{n} close prices, here $n=5050$. When DY compute the log-returns dataset \emph{logReturns} starting from \emph{closePrice}, they obtain the following time series:
$$
logReturns = \{r_t | t \in [1, \ n-\tau] \}
$$  
where $r_t = log \frac{P_{t+\tau}}{P_t}, \ \forall t \in [1, \ n-\tau]$.\\
In our example, we would have
\begin{eqnarray*}
logReturns & = & \{ r_1, r_2, ..., r_{n-\tau} \} \\
& = & \{ log\frac{P_{1+\tau}}{P_1}, log\frac{P_{2+\tau}}{P_2}, ..., log\frac{P_n}{P_{n-\tau}}  \} \\
& = & \{ log\frac{P_6}{P_1}, log\frac{P_7}{P_2}, ..., log\frac{P_{5050}}{P_{5045}}  \} \\
\end{eqnarray*}
Thus, they obtain a single dataset of $n-\tau$ log-returns. We believe this way of computing the log-returns time series is biased, because it ``re-uses" the data. Indeed, let us assume that a crash occurs at time $t^*$. Then they will take into account this specific event $\tau$ times exactly in their dataset in log-returns $\{ r_{t^*-\tau}, r_{t^*-\tau+1}, ...,r_{t^*-1 } \}$. Working on this time series will definitely fatten the tails, since every shock in the original price time series will appear exactly $\tau$ times in the log-returns. Hence, the resulting log-returns time series is not an accurate representation of the market moves anymore; this gives a decisive advantage to their model compared with the Gaussian, since they train their parameters to fit specifically the fat tails. 

Finally, DY do not statistically test the goodness-of-fit of their formula, but rather rely exclusively on plots that truly look good.\\

\section{Goodness-of-fit tests}
\label{statisticalTests}

For these reasons, we decided to keep the outliers in our dataset and to use multiple non-overlapping log-return time series. We perform our tests on the Dow-Jones Industrial Average, from January 04, 1982 to December 31, 2001. From the raw price time series, we derive $\tau$ log-return time series $logReturns_j$ ($j \in [1, \tau]$) of cardinality $m=[\frac{n}{\tau}]$, called ``paths", instead of a single log-return time series $logReturn$ of cardinality $n-\tau$. This method sounds more reasonable for very different reasons: firstly, when economists or traders talk about weekly returns ($\tau=5$ days, holidays excluded), they mean returns from Mondays to Mondays (for instance), and not from Monday to Monday, Tuesday to Tuesday, etc. Secondly, this method gives a better view of the real market, since tails are not artificially fattened. And finally, this method will enable us to assess the statistical accuracy of the estimators we will compute.\\
We can now perform our goodness-of-fit tests on our different models: the Gaussian (\emph{normPDF}), the curve resulting from the DY formula (\emph{draguPDF}), and a Neural Network\footnote{Even if \emph{draguPDF} is supposed to fit the empirical distribution, \emph{empPDF}, better than \emph{normPDF}, we want to compare it with the best fit possible, the one obtained with a Neural Network. This Neural Network must be as simple as possible, but should fit the main characteristics of the empirical time series, fat tail and kurtosis. The structure chosen was the following: it is a feed-forward back-propagation network, with a five node hidden layer and a single node output layer. The transfer functions are respectively $tansig$ and $purelin$, where $tansig(n) = \frac{2}{1+e^{-2*n}}-1$ and $purelin(n)=n$. This structure appears to be a good trade-off between the complexity and the goodness of fit.} intentionally overfitting the data, that will be used as a benchmark (\emph{nnPDF}).

\subsection{Kolmogorov-Smirnov Statistic}

DY claim that their model fits the empirical data of the concerned dataset better than the Gaussian for any time lag. To check this, we use the Kolmogorov-Smirnov Statistic, based on the maximal discrepancy between the expected and the observed cumulative distributions, for any log-return $x$. This statistic is suitable for testing only a simple hypothesis, for instance a Gaussian with known parameters $\mu$ and $\sigma$, but not a composite hypothesis (a class of Gaussians, or a Gaussian with $\mu$ and $\sigma$ derivated from the sample dataset being tested). Unfortunately, whatever the model, we always derive the parameters ($\mu$ and $\sigma$ for \emph{normPDF}, $\gamma$, $\theta$, $k$ and $\mu$ for \emph{draguPDF}, the weights and biases for \emph{nnPDF}) from the initial dataset. By performing this test with parameters derived from the dataset being tested, we expect the statistic to be large enough to reject the simple hypothesis, and \emph{a fortiori} the composite hypothesis (\cite{Bree}). But if the value of the statistic $Z$ is small enough to accept the simple hypothesis, it does not mean that we can accept the composite hypothesis.\\

\begin{figure}[t]
\centering
\vspace{2.5mm}
\centerline{\includegraphics[width=.75 \columnwidth]{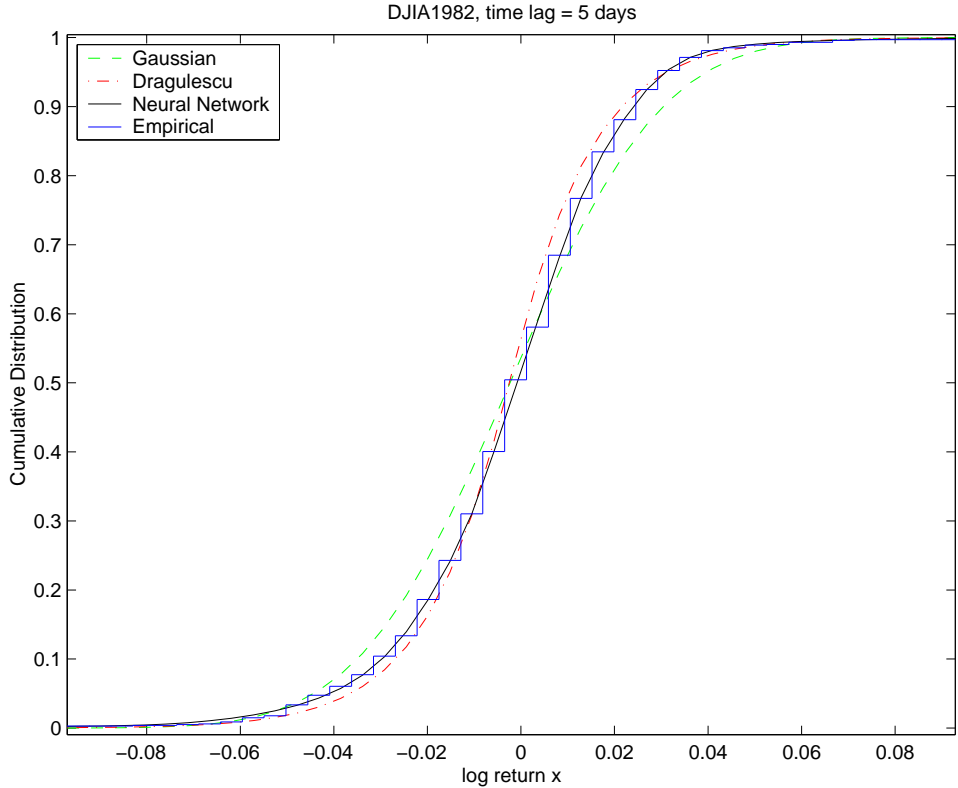}}
\vspace{2mm}
\caption{Cumulative Density Functions for the different models, $\tau=5$ days.}
\label{fig:fig-CDF}
\end{figure}

{\bfseries Methodology -} For each time lag, we compute the log-returns dataset, and we divide it into paths. For each path and each model, we build the empirical cumulative density function \emph{empCDF} and the expected CDF \emph{modelCDF} (\emph{normCDF}, \emph{draguCDF} or \emph{nnCDF}), and we compute the KS-statistic $Z$ (see Figure \ref{fig:fig-CDF} for an example). We present in Tables \ref{tab:normKolmo}, \ref{tab:draguKolmo} and \ref{tab:nnKolmo} the mean $\tilde{Z}$ and standard deviation $\sigma_Z$ of $Z$ over the different paths, and the associated p-value\footnote{The p-value is the probability of observing the given sample result under the assumption that the null hypothesis (the tested model) is true. If the p-value is less than the level of significance $\alpha$, then you reject the null hypothesis. For example, if $\alpha = 0.05$ and the p-value is 0.03, then you reject the null hypothesis.} interval $p(\tilde{Z}+\sigma_Z) \leq p(\tilde{Z}) \leq p(\tilde{Z}-\sigma_Z)$.\\

{\bfseries Results -} Firstly, we observe an important variance, over the different paths, in the statistic $Z$: the standard deviation $\sigma_Z$ is not negligible in comparison with the mean $\tilde{Z}$. This is an evidence that the paths are not equivalent, which legitimates, \emph{a posteriori}, our sampling method. It comes from the high heterogeneity of the dataset, which makes our tests less robust. But any test performed on this heterogeneous dataset would suffer from the same problem. Even though this apparent lack of consistency prevents us from drawing any strong and global conclusion, the knowledge of the mean and standard deviation $\tilde{Z} \pm \sigma_Z$ provides us with a fair overview of the statistic $Z$.\\
On plots, the DY formula seems to fit the empirical cumulative distribution better than the Gaussian. But in fact, on average, both models are rejected for high  frequencies (for $\tau=1 \ \mbox{and}\  5$ days) at the 0.01 level of significance. Even the Neural Network is rejected for a one day time lag. This rejection of the three models may come from the fact that this test is based on the maximum discrepancy between the empirical and the theoretical cumulative distributions, \emph{for any} $x$. To pass this test, a model must fit the observed data sufficiently well everywhere, i.e. in the tails (problem of fat tails) and in the middle (problem of high kurtosis for high frequencies) of the distribution.\\
We point out that even if the DY formula is rejected for a one day time lag, the statistic $Z$ is smaller than the Gaussian one (0.109 vs 0.131), which is an indication that the model fits the data a bit better. For other time lags, the p-value are equivalent: both models are systematically rejected for 5 days ($p \ll 0.01$), sometimes rejected for 20 days ($p(\tilde{Z}+\sigma_Z) \leq 0.01$, but $p(\tilde{Z}-\sigma_Z) \geq 0.05$) and never rejected for higher frequencies ($p \gg 0.05$). For medium and low frequencies, the fact that the simple hypothesis is not rejected does not guarantee that the composite hypothesis can be accepted.\\ 

{\bfseries Conclusion -} The Kolmogorov-Smirnov goodness-of-fit test rejects both the Gaussian and the DY formula for high frequencies ($\tau=1 \ \mbox{and}\  5$ days). For medium and low frequencies, we cannot come to a firm conclusion because of the theoretical limits of this test. To continue with the investigation, we need a more powerful statistical test that can be used even when the parameters of the model are derived from the dataset being tested. The $\chi^2$ statistic is suitable in those conditions.

\begin{table}[p]
\begin{center}
\begin{tabular}{|c|c|c|}
\hline
time lag & $\tilde{Z} \pm \sigma_{Z}$ & p-values \\ 
\hline
\begin{tabular}{r}
	 1 \\
	 5 \\
	 20 \\
	 40 \\
	 80 \\
	 100 \\
	 200 \\
	 250 \\
	\end{tabular} &
\begin{tabular}{l}
       0.131         \\
       0.081 $\pm$0.020 \\
       0.089 $\pm$0.013 \\
       0.104 $\pm$0.020 \\
       0.113 $\pm$0.021 \\
       0.113 $\pm$0.021 \\
       0.148 $\pm$0.038 \\
       0.170 $\pm$0.047 \\ 
	\end{tabular} &
\begin{tabular}{r@{}lrr@{}lrr@{}l}
	&&& 2&.93e-75 &&& \\
	1&.75e-09 & $\leq$ & 2&.98e-06 & $\leq$ & 9&.88e-04\\
	9&.51e-03 & $\leq$ & 0&.036    & $\leq$ & 0&.112\\
	0&.038    & $\leq$ & 0&.122    & $\leq$ & 0&.321\\
	0&.199    & $\leq$ & 0&.385    & $\leq$ & 0&.649\\
	0&.322    & $\leq$ & 0&.533    & $\leq$ & 0&.778\\
	0&.339    & $\leq$ & 0&.630    & $\leq$ & 0&.917\\
	0&.291    & $\leq$ & 0&.598    & $\leq$ & 0&.919\\
	\end{tabular} \\
\hline
\end{tabular}
\end{center}
\caption{KS-Test on the Gaussian.}
\label{tab:normKolmo}
\end{table}

\begin{table}[p]
\begin{center}
\begin{tabular}{|c|c|c|}
\hline
time lag & $\tilde{Z} \pm \sigma_{Z}$ & p-values \\ 
\hline
\begin{tabular}{r}
	 1 \\
	 5 \\
	 20 \\
	 40 \\
	 80 \\
	 100 \\
	 200 \\
	 250 \\
	\end{tabular} &
\begin{tabular}{l}
	0.109 \\        
	0.087 $\pm$0.019  \\
	0.089 $\pm$0.014 \\
	0.094 $\pm$0.010 \\
	0.116 $\pm$0.018 \\
	0.128 $\pm$0.019 \\
	0.163 $\pm$0.048 \\
	0.186 $\pm$0.046 \\
	\end{tabular} &
\begin{tabular}{r@{}lrr@{}lrr@{}l}
	&&& 1&.2e-53  &&& \\
	2&.08e-10 & $\leq$ & 3&.64e-07 & $\leq$ & 1&.48e-04\\
	8&.75e-03 & $\leq$ & 0&.033    & $\leq$ & 0&.104\\
	0&.125    & $\leq$ & 0&.211    & $\leq$ & 0&.337\\
	0&.197    & $\leq$ & 0&.355    & $\leq$ & 0&.576\\
	0&.215    & $\leq$ & 0&.372    & $\leq$ & 0&.585\\
	0&.209    & $\leq$ & 0&.512    & $\leq$ & 0&.893\\
	0&.224    & $\leq$ & 0&.481    & $\leq$ & 0&.816\\
	\end{tabular} \\
\hline
\end{tabular}
\end{center}
\caption{KS-Test on the DY formula.}
\label{tab:draguKolmo}
\end{table}

\begin{table}[p]
\begin{center}
\begin{tabular}{|c|c|c|}
\hline
time lag & $\tilde{Z} \pm \sigma_{Z}$ & p-values \\  
\hline
\begin{tabular}{r}
	 1 \\
	 5 \\
	 20 \\
	 40 \\
	 80 \\
	 100 \\
	 200 \\
	 250 \\
	\end{tabular} &
\begin{tabular}{l}
	0.106  \\
	0.048 $\pm$0.014 \\
	0.047 $\pm$0.009 \\
	0.071 $\pm$0.059 \\
	0.075 $\pm$0.061 \\
	0.076 $\pm$0.039 \\
	0.116 $\pm$0.034 \\
	0.137 $\pm$0.041 \\
	\end{tabular} &
\begin{tabular}{r@{}lrr@{}lrr@{}l}
	&&& 3&.27e-42 &&& \\
	1&.64e-03 & $\leq$ & 0&.026 & $\leq$ & 0&.204\\
	0&.430 & $\leq$ & 0&.615 & $\leq$ & 0&.778\\
	0&.153 & $\leq$ & 0&.746 & $\leq$ & 1&\\
	0&.729 & $\leq$ & 0&.919 & $\leq$ & 0&.995\\
	0&.532 & $\leq$ & 0&.871 & $\leq$ & 0&.999\\
	0&.126 & $\leq$ & 0&.453 & $\leq$ & 0&.932\\
	0&.190 & $\leq$ & 0&.502 & $\leq$ & 0&.904\\
	\end{tabular} \\
\hline
\end{tabular}
\end{center}
\caption{KS-Test on the Neural Network.}
\label{tab:nnKolmo}
\end{table}

\subsection{$\chi^2$ Statistic}

The $\chi^2$ goodness-of-fit test, based on binned data, is a powerful statistical tool to test if an empirical sample comes from a given distribution. Contrary to the Kolmogorov-Smirnov test, it is designed to evaluate a composite hypothesis, i.e. the parameters of the model can be derivated from the empirical dataset tested. This test is a good trade-off between the goodness-of-fit of a model (the better fit, the smaller the $\chi^2$ statistic) and its complexity (the more complex, the smaller p-value). Indeed, even if a model fits the empirical data very well, a too large complexity may penalise its p-value, so that it can still be rejected. Finally, to be meaningful, this test must be performed using relatively large bins, and a critical value of 5 expected observations per bin is regarded as a minimum.\\  

{\bfseries Methodology -} If we perform this test with equal size bins, then the fats tails will be trimmed (there are less than 5 expected log-returns per bin in the tails) and will not participate in the value of the statistic, making the test irelevant. Instead, we split the log-return axis into \emph{equal expected frequency bins}, so that all of the log-returns participate in the value of the statistic. We use an expected frequency of 5 log-returns per bin. Unfortunately, this test cannot be performed for large time lags, because of the lack of data. Indeed, for the Dow-Jones index from 1982 to 2001 for instance, we have initially around 5000 close prices, which means that for a time lag of 250 days, each path will have only about 20 log-returns. In those conditions, because of the critical value of 5 log-returns per bin, we will have at best 4 bins, which is too small to perform a relevant test.\\

{\bfseries Results -}  We present our results of the $\chi^2$ goodness-of-fit test in Tables \ref{tab:normChi}, \ref{tab:draguChi} and \ref{tab:nnChi}. The degree of freedom is given by $df=noBins-1-m$, where $m$ is the number of parameters of the model ($m=2$ for the Gaussian, $m=4$ in the DY formula, and $m=11$ for the Neural Networks if we count the weights and the biases). For large time lags, $df$ becomes smaller and smaller because $noBins$ decreases, as explained above.\\
Concerning the Neural Network, we cannot perform this test for time lags higher than 40 days, or else the degree of freedom decreases to zero. This is due to the relatively high number of parameters ($m=11$). With a structure even more complicated, we could not have performed the test at all, except for high frequencies.\\
First we notice that the Neural Network's $\chi^2$ statistic is slightly smaller than the DY's one, itself smaller than the Gaussian's one, for all paths with a time lag from $\tau=1$ to $\tau=80$ days. It means that the Neural Networks fits empirical data better than the DY formula, which itself has a better fit than the Gaussian. But there is a price to pay, in terms of complexity: due to too many parameters (and then a lower degree of freedom), the p-value of the Neural Networks and DY formula are not systematically higher than the p-value of the Gaussian. And it is precisely the p-value that is used to accept or reject a model, not directly the $\chi^2$ statistic.\\
If we look at the p-value in detail, we observe that
\begin{itemize}
\item For $\tau=1$, the three models are rejected at a 0.05 level of significance
\item For $\tau=5$, only the Neural Network is systematically accepted. The Gaussian and DY formula are only accepted in the best situation ($p(\tilde{\chi^2}) < 0.05 < p(\tilde{\chi^2}-\sigma_{\chi^2})$)
\item For $\tau=20$, the three models are accepted and the DY formula is better than the Neural Network and the Gaussian
\item For $\tau=40$ and $\tau=80$, the DY formula is still accepted, but its p-value is smaller than the one of the Gaussian 
\end{itemize}

{\bfseries Conclusion -} Thanks to the $\chi^2$ goodness-of-fit test, we can assert that the DY formula fits empirical data slightly better than the Gaussian, for high and medium frequencies. Nevertheless, both models are rejected for high frequencies ($\tau=1 \ \mbox{and}\  5$ days), at a 0.05 level of significance. In this sense, these results are consistent with the Kolmogorov-Smirnov goodness-of-fit test.\\
We also observe a clear shift in the goodness-of-fit of the models around $\tau=40$ days: the probability of accepting the Gaussian becomes larger than the probability to accept the DY formula (and even the Neural Network) due to the lower complexity of the Gaussian (two parameters instead of four and eleven respectively).\\
To put it in a nutshell, using a complex model, such as the DY formula or a Neural Network, is only worthwhile for $\tau=1, 5 \ \mbox{and} \ 20$ days. For lower frequencies ($\tau \geq 40$ days), the Gaussian is preferable because it is simpler. Given that for these frequencies, we had observed neither fat tails nor kurtosis in the empirical datasets, the Gaussian represents the best trade-off between goodness-of-fit and complexity.\\

\begin{table}[p]
\begin{center}
\begin{tabular}{|c|c|c|c|}
\hline
time lag & $\tilde{\chi^2} \pm \sigma{\chi^2}$ & df & p-values \\ 
\hline
\begin{tabular}{r}
	1 \\
	5 \\
	20 \\
	40 \\
	80 \\
	\end{tabular} &
\begin{tabular}{r@{}ll} % norm
	1790&& \\
	255&& $\pm$30\\
	61&& $\pm$12\\
	29&.1 &$\pm$7.0\\
	10&.4 &$\pm$4.6\\
	\end{tabular} & 
\begin{tabular}{r}
	1010 \\
	198 \\
	47 \\
	22 \\
	9 \\
	\end{tabular} & 
\begin{tabular}{r@{}lrr@{}lrr@{}l}
	&&& 6&.29e-11 &&&\\
	5&.38e-05 & $\leq$ & 4&.07e-03 & $\leq$ & 0&.0931\\
	7&.99e-03 & $\leq$ & 0&.0819 & $\leq$ & 0&.409\\
	0&.0295 & $\leq$ & 0&.141 & $\leq$ & 0&.451\\
	0&.0915 & $\leq$ & 0&.32 & $\leq$ & 0&.76\\
	\end{tabular} \\
\hline
\end{tabular}
\end{center}
\caption{$\chi^2$ Test on the Gaussian.}
\label{tab:normChi}
\end{table}

\begin{table}[p]
\begin{center}
\begin{tabular}{|c|c|c|c|}
\hline
time lag & $\tilde{\chi^2} \pm \sigma{\chi^2}$ & df & p-values \\ 
\hline
\begin{tabular}{r}
	1 \\
	5 \\
	20 \\
	40 \\
	80 \\
	\end{tabular} &
\begin{tabular}{r@{}ll} % dragu
	\tt 1420 \\
	\tt 244& & $\pm$26 \\
	\tt 48&.5 & $\pm$11.5 \\
	\tt 27&.3 & $\pm$6.1 \\
	\tt 9&.7 & $\pm$4.4 \\
	\end{tabular} & 
\begin{tabular}{r}
	\tt 1000 \\
	\tt 196  \\
	\tt 45 \\
	\tt 20  \\
	\tt 7 \\
	\end{tabular} & 
\begin{tabular}{r@{}lrr@{}lrr@{}l}
	&&& 1&.16e-04 &&& \\
	3&32e-04 & $\leq$ & 0&.0108 & $\leq$ & 0&.133\\
	0&.0663 & $\leq$ & 0&.333 & $\leq$ & 0&.796\\
	0&.0301 & $\leq$ & 0&.126 & $\leq$ & 0&.385\\
	0&.049 & $\leq$ & 0&.206 & $\leq$ & 0&.624\\
	\end{tabular} \\
\hline
\end{tabular}
\end{center}
\caption{$\chi^2$ Test on the DY formula.}
\label{tab:draguChi}
\end{table}

\begin{table}[p]
\begin{center}
\begin{tabular}{|c|c|c|c|}
\hline
time lag & $\tilde{\chi^2} \pm \sigma{\chi^2}$ & df & p-values \\ 
\hline
\begin{tabular}{r}
	1 \\
	5 \\
	20 \\
	40 \\
	80 \\
	\end{tabular} &
\begin{tabular}{r@{}ll} % nn
	\tt 2230 \\
	\tt 232& & $\pm$38 \\
	\tt 45&.9 & $\pm$11.1 \\
	\tt 21&.5 & $\pm$6.3 \\
	\tt 7&.6 & $\pm$6.3 \\
	\end{tabular} & 
\begin{tabular}{r}
	\tt 997 \\
	\tt 189 \\
	\tt 38 \\
	\tt 13 \\
	\tt 0 \\
	\end{tabular} & 
\begin{tabular}{r@{}lrr@{}lrr@{}l}
	&&& 0&.0839 &&& \\
	0&.0559 & $\leq$ & 0&.346 & $\leq$ & 0&.817\\
	0&.0473 & $\leq$ & 0&.25 & $\leq$ & 0&.688\\
	0&.0057 & $\leq$ & 0&.0552 & $\leq$ & 0&.333\\
	&NaN & $\leq$ & &NaN & $\leq$ & &NaN\\
	\end{tabular} \\
\hline
\end{tabular}
\end{center}
\caption{$\chi^2$ Test on the Neural Network.}
\label{tab:nnChi}
\end{table}

\newpage
   
\section{Conclusion}
\label{conclusion}

The DY analytical formula is quite exciting, and would validate the Heston model, if the resulting expected distribution was similar to the empirical one. Our method of pre-processing the data is different from the one suggested by Dr\u{a}gulescu and Yakovenko, and, we believe, more truly captures the market. We find that the DY formula consistently outperforms the Gaussian in terms of best fit (the error between the model and the observed data), but a higher complexity is the price to pay: the number of parameters is greater in the Heston model, which enables the fitted \emph{pdf} to exhibit kurtosis and fat tails when the empirical dataset does, but also penalizes it when it doesn't. Hence, in terms of goodness-of-fit (the trade-off between best fit and complexity), the Gaussian is preferable for low frequencies ($40 \leq t \leq 250$ days), since the empirical dataset is then quite smooth and does not exhibit kurtosis. For medium frequencies ($t = 20$ days), both models are accepted at a .05 confidence level. Finally, for high frequencies, although it performs better than the Gaussian, the DY formula is still rejected ($t = 1$ and $5$ days), mainly because of the extremely high kurtosis of the observed data.

\end{document}